\documentclass[12pt]{iopart}
\usepackage{graphicx}
\usepackage{epstopdf}
\usepackage{breqn}
\usepackage{color}
\usepackage{amsmath,amsthm, amssymb, latexsym}

\renewcommand{\figurename}{Figure}

\begin{document}

\title[Ponderomotive force driven mechanism for electrostatic wave excitation]{Ponderomotive force driven mechanism for electrostatic wave excitation and energy absorption of Electromagnetic waves  in overdense magnetized plasma}

\author{Laxman Prasad Goswami$^{*1}$, Srimanta Maity$^{*2}$, Devshree Mandal$^{\dagger 3}$, Ayushi Vashistha$^{\dagger 4}$ and Amita Das$^{*5}$}

\address{$^{*}$Department of Physics, Indian Institute of Technology Delhi, Hauz Khas, New Delhi-110016, India\\ $^{\dagger}$Institute for Plasma Research, HBNI, Bhat, Gandhinagar-382428, India \\}
\ead{$^{1}$goswami.laxman@gmail.com, $^2$srimantamaity96@gmail.com, $^3$ayushivashistha@gmail.com, $^4$devshreemandal@gmail.com, $^5$amita@iitd.ac.in}
\vspace{10pt}
\begin{indented}
\item[]April 2021
\end{indented}

\begin{abstract}
	The excitation of electrostatic waves in plasma by laser electromagnetic pulse is important as it provides a scheme by which the power from the laser electromagnetic (EM) field can be transferred into the plasma medium. The paper presents a fundamentally new ponderomotive pressure-driven mechanism of excitation of electrostatic waves in an overdense magnetized plasma by a finite laser pulse. 
	Particle-in-cell (PIC) simulations using the EPOCH-4.17.10 framework have been utilized for the study of a finite laser pulse interacting with a  
	magnetized overdense plasma medium. The external magnetic field is chosen to be aligned parallel to the laser propagation direction. 
	In this geometry, the electromagnetic wave propagation inside the plasma is identified as whistler or R and L waves. The group velocity of these waves 
	being different, a clear spatial separation of the R and L 
	pulses are visible.   In addition, excitation of electrostatic perturbation associated with the EM pulses propagating inside the plasma is also observed.  These electrostatic perturbations are important as they couple laser energy to the plasma medium.   The  excitation of electrostatic oscillations are  understood here by a fundamentally new mechanism of 
	charge separation created by the  difference between the ponderomotive force  (of the electromagnetic pulse) felt by the two plasma species, viz., the  
	electrons and the ions in a magnetized plasma. 
\end{abstract}

\section{Introduction}
It is well known that the interaction of  electromagnetic wave (EM) with  magnetized plasma is considerably rich 
compared to un-magnetized plasma \cite{MagnetizedPlasma1, MagnetizedPlasma2, MagnetizedPlasma3, MagnetizedPlasma4}. 
This regime, however, has not yet  been explored in the laboratory laser-plasma experiments due to the magnitude of the magnetic field required to make even the lighter 
electron species of  
the plasma magnetized at the laser frequency. With the recent technological advancements, magnetic fields of the order of kilo Tesla have now been achieved \cite{1200TMagnetic}. This is typically a factor of $10$   smaller than the required magnetic field strength for making the lighter electron species magnetized at the $CO_2$ laser frequency. {Recently a theoretical mechanism has been proposed for the generation of much higher magnetic fields \cite{HighMagneticField}.}
It appears, therefore, that there is a good likelihood for the technological advancements to catch up in near future to have the regime of magnetized electron response accessible in laser-plasma interaction experiments.  The other regime where even the heavier ions become magnetized appear considerably difficult to achieve in laboratory laser-plasma experiments in near future. Both, these regimes, however, are of importance in astrophysical scenarios where an EM wave can readily encounter a strongly magnetized plasma \cite{AstrophysicalMagneticField, AstrophysicalWaves}. It is, therefore, important to understand the underlying physics of 
the EM wave interaction with a strongly magnetized plasma in detail. 
It is important to note that Particle - In - Cell (PIC) simulations have in recent decades developed a sophistication to capture the physics of laser  (or EM wave ) plasma interaction phenomena in their pristine glory \cite{PIC1, PIC2}. We have, therefore,  chosen to carry out PIC studies to investigate the interaction of laser with a magnetized plasma which cannot be explored as yet in laboratory experiments. {There exists many familiar schemes that facilitate the laser energy absorption in the plasma \cite{Absorption1, Absorption2}. For instance, $\vec{J}\times \vec{B}$, resonance absorption, Brunel heating schemes etc. \cite{Absorption3, Absorption4, Absorption5, Absorption6, Absorption7, Absorption8, Absorption9}. }  

It has been shown in earlier studies \cite{MagnetizedPlasma4, NewMechanism1, NewMechanism3} concerning the magnetized laser plasma interaction that when the external magnetic field is directed normal to both the propagation direction and the direction of laser electric field  (X mode geometry) excitation of electrostatic lower Hybrid wave are observed as a result of the difference between the $\vec{E} \times \vec{B}$ drifts of ions and electrons. This electrostatic coupling was responsible for a  new mechanism of laser energy absorption. In the O mode geometry though the EM wave 
is unable to penetrate the overdense plasma, it generates  harmonics that propagate inside the plasma 
if appropriate conditions are satisfied \cite{harmonic}.  
In this work, we restrict ourselves  to a geometry for which the external magnetic field is directed along the propagation direction 
of the laser pulse.

The paper has been organized as follows. Section \ref{SimulationDetails} contains the simulation details. In section \ref{Observation} we present the observations wherein we observe an electrostatic excitation in the plasma trailing the transmitted electromagnetic signal inside the plasma. 
The results are analyzed in section \ref{ElectrostaticExcitation}.   We  identify the mechanism 
of the generation of electrostatic fluctuations in the plasma to the difference in the ponderomotive force of the laser field felt by the two species (electrons and ions) in the presence of an external magnetic field \cite{Ponderomotive1, Ponderomotive2, Ponderomotive3, Ponderomotive4, Ponderomotive5}. 
Section \ref{conclusion} provides the summary and conclusion.


\section{Simulation details}
\label{SimulationDetails}

\begin{figure*}
	\centering
	\includegraphics[width=6.5in]{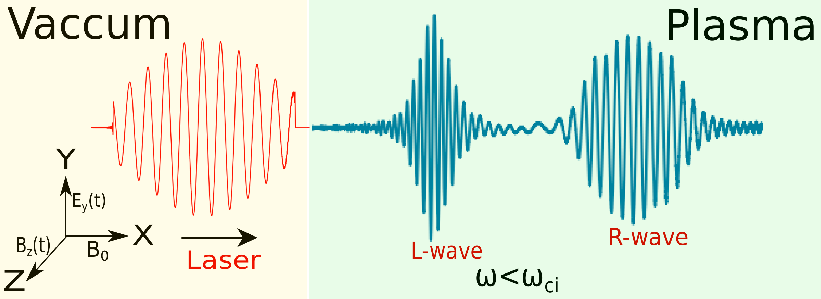}
	\caption{Schematic (not to scale) showing laser energy being coupled into plasma via excitation of L and R waves in the system. The external magnetic field $B_0$ is along laser propagation direction ($\hat{x}$). The laser electric field is in along $\hat{y}$ while the laser magnetic field is in the $\hat{z}$ direction. The laser frequency $\omega$ is taken less tha  the ion cyclotron frequency $\omega_{ci}$.}
	\label{fig:schematic}
\end{figure*}

We have employed  EPOCH-4.17.10  for carrying out  1-D particle-in-cell (PIC) simulations to study the interaction of a laser with plasma in the presence of an external magnetic field \cite{EPOCH}. 
The magnetic field is directed along the laser propagation direction.  The schematic of the simulation geometry (not to scale)  has been shown in 
Fig.\textcolor{blue}{\ref{fig:schematic}}. The laser is propagating along  $\hat{x}$ direction and the  applied external magnetic field is also along  $\hat{x}$. The  electric field of laser is directed along $\hat{y}$. A 1-D simulation box with dimension $L_{x} = 1000 \mu m$ has been chosen. The number of particles per cell are taken to be $200$. The plasma boundary starts from $x = 0$. There is vacuum between $x = -150$ to $ 0$ $\mu m$. 
At $x=0$, there is a sharp  plasma vacuum interface. 
The spatial resolution is taken as $30$ cells per electron skin depth and it  corresponds to the grid size $\Delta x = 0.01 \mu m$. The laser is incident on the plasma target from left side. We consider a short-pulse laser of wavelength $\lambda_{l} = 9.42$ $\mu m$ (frequency $\omega_l = 0.2\times 10^{15} rad/s$), 
which corresponds to $CO_2$ laser. The choice of long wavelength $CO_2$ laser reduces the requirements of the externally applied magnetic field for allowing magnetized response of plasma.  The laser profile is Gaussian with the peak intensity of $I = 3.5 \times 10^{19} Wm^{-2}$ . Boundary conditions are taken as absorbing in the longitudinal direction. The number density of the plasma is taken to be $n_{0} = 3 \times 10^{26} m^{-3}$ for which the electron plasma frequency is equal to $10^{15} rad/s$. Thus, the plasma is overdense for the incident laser pulse. 

The dynamics of both electrons and ions has been followed. To reduce the computational time, we carried out the simulations at a reduced mass of ions, which is taken to be $25$ (or $100$ for a few selected runs) times heavier than electrons (i.e. $m_{i} = 25m_{e}$, where $m_{i}$ and $m_{e}$ represent the rest mass of the ion and electron species respectively). The external magnetic field is chosen in such a way that for simulations in case(A) the condition $\omega_{ci} < \omega_l < \omega _{ce}$ is satisfied. To satisfy this condition the value of magnetic field has been taken to be $2.5 m_{e} c \omega _{pe} e^{-1}$. For case (B) we have 
$ \omega_l < \omega_{ci} < \omega _{ce}$ which corresponds to magnetic field of $10 m_{e} c \omega _{pe} e^{-1}$.

In this work, we  identify a   new mechanism of electrostatic wave excitation in the plasma by laser which depends on the difference between the ponderomotive force 
of the EM radiation felt by the electron and ions in the presence of an external magnetic field. 
To distinguish it from all other mechanisms we choose a specific simulation configuration. 
The laser is chosen to be incident on the sharp plasma surface in the normal direction which ensures the absence of resonance and vacuum heating processes. 
Also, the role of $\vec{J} \times \vec{B}$ electron heating is made negligible by choosing the laser intensity in the non-relativistic domain of $ a_{0} (= eE/m_e \omega_l c) <1$ ($a_0 = 0.6$ ). The difference in $\vec{E} \times \vec{B}$ drift velocities felt by the electron and ions in the presence of external magnetic field reported in \cite{MagnetizedPlasma4, NewMechanism1} will also not be operative in the longitudinal direction as the applied magnetic field is also directed  along $\hat{x}$. We have provided the chosen values of the laser and plasma simulation parameters in Table-\textcolor{blue}{\ref{simulationtable}}. 

\begin{table}
	\caption{Values of simulation parameters in normalized and corresponding standard units for $m_i = 25$}
	\label{simulationtable}
	\footnotesize
	\begin{center}
		\begin{tabular}{|c||c||c|}
			\hline		
			\color{red}Parameters&\color{red}Values in SI unit&\color{red}Values in SI unit\\
			&\color{black}Case A &\color{black}Case B\\
			\hline
			\hline	
			\multicolumn{3}{|c|}{\color{blue}Plasma Parameters} \\
			\hline
			$n_0$&$3\times10^{20} cm^{-3}$&$3\times10^{20} cm^{-3}$\\
			\hline
			$\omega_{pe}$&$10^{15} rad/s$&$10^{15} rad/s$\\
			\hline
			$\omega_{pi}$ ($M/m = 25$)&$0.2\times10^{15} rad/s$&$0.2\times10^{15} rad/s$\\
			\hline
			\multicolumn{3}{|c|}{\color{blue}Laser Parameters} \\
			\hline
			$\omega_{l}$&$0.2\times10^{15} rad/s$&$0.2\times10^{15} rad/s$\\
			\hline
			$\lambda_{l}$&$9.42 \mu m$&$9.42 \mu m$\\
			\hline
			Intensity&$3.5\times 10^{19} W/m^2$&$3.5\times 10^{19} W/m^2$\\
			\hline
			\multicolumn{3}{|c|}{\color{blue}External Magnetic Field Parameters} \\
			\hline
			$B_{0}$&$2.5 m_{e} c \omega _{pe} e^{-1}$&$10 m_{e} c \omega _{pe} e^{-1}$\\		
			\hline
			$\omega_{ce}$&$2.5\times10^{15} rad/s$&$10\times10^{15} rad/s$\\
			\hline
			$\omega_{ci}$&$0.1\times10^{15} rad/s$&$0.4\times10^{15} rad/s$\\
			\hline
		\end{tabular}\\	
	\end{center}
\end{table}

\normalsize

\section{Observations}
\label{Observation}
We have considered two cases in our simulation. One for which the magnetic field is such that only the lighter electron species is magnetized at the laser frequency and another where both the electron and ion species are magnetized. We refer to these two cases as case (A) and case (B) respectively. 
Since the applied magnetic field is along the direction of laser propagation,  for case (A) one observes the EM wave to get converted to the whistler (R) wave as it enters the plasma medium. In case(B) both L and R waves form and are seen to get separated spatially as their group speeds differ in the plasma medium.  
We also observe the formation of electrostatic disturbances in the plasma medium.  

\subsection{Case (A)}
\begin{figure*}
	\centering
	\includegraphics[width=6.2in]{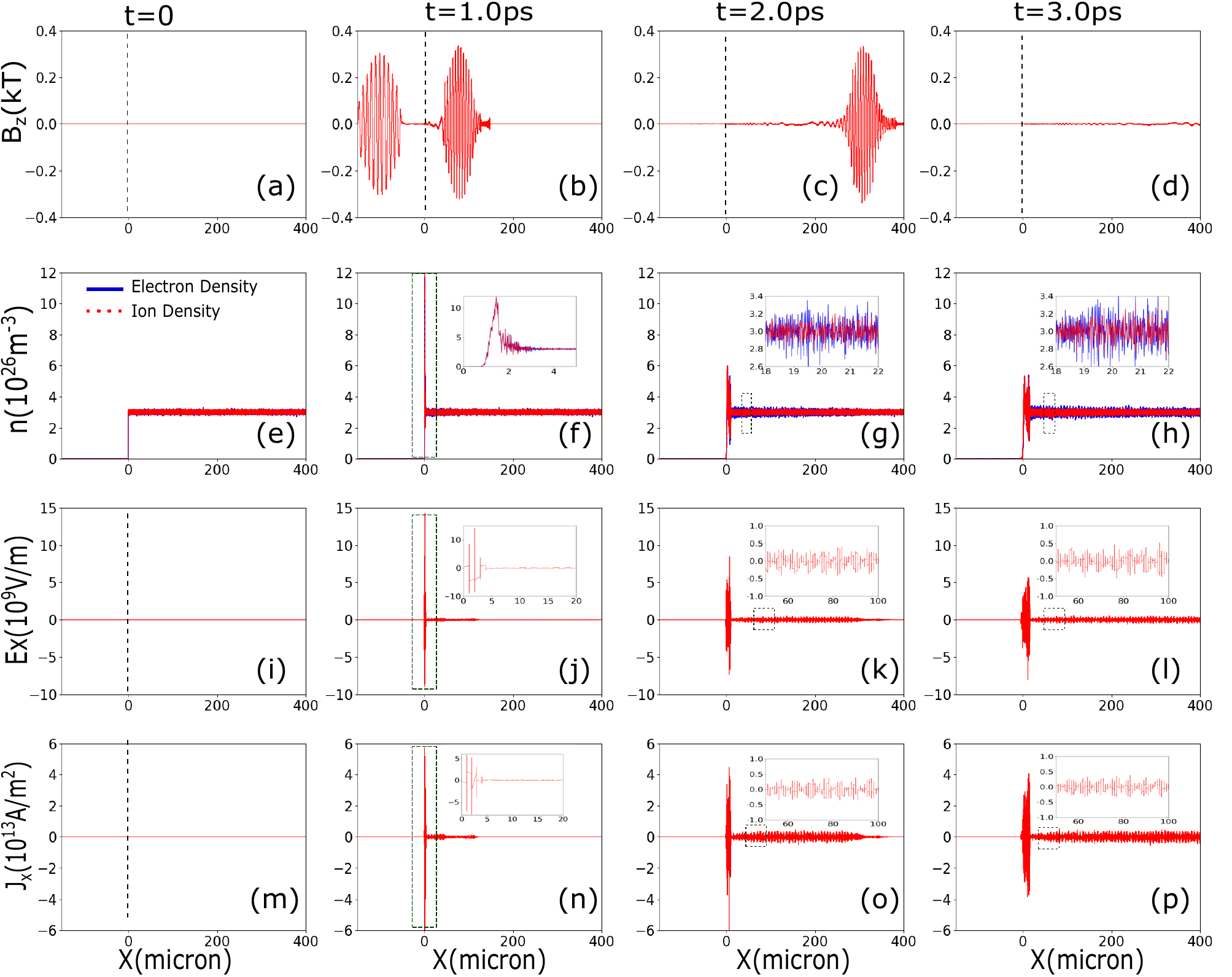}
	\caption{Time evolution of spatial variation of $z$-component of magnetic field $B_z$,  perturbation of electron and ions density ($n$), $x$-component of electric field $E_x$ and $x$-component of total current $J_x$ with respect to $\hat{x}$ for case (A), showing that as the laser interacts with magnetized plasma, there is finite magnitude of perturbed $B_z, E_x$ and $J_x$ inside the plasma. We can infer from these plots that a part of laser energy has been coupled into plasma.}
	\label{fig:WhistlerObervation}
\end{figure*}

In this case, the laser frequency satisfies the condition ($\omega_{ci}<\omega_{l}<\omega_{ce}$) such that only the lighter electron species are magnetized.  In Fig. \textcolor{blue}{\ref{fig:WhistlerObervation}} we show the evolution of various fields as a function of $x$   at different  times 
(viz., $t = 0$, $t= 1ps$, $t = 2.0ps$ and $t = 4 ps$). The laser  magnetic field, shown in the  
Fig. \textcolor{blue}{\ref{fig:WhistlerObervation}}(a), (b), (c) and (d) shows the reflection and transmission of the laser field from the plasma surface. The incident laser is plane-polarized with its electric field directed along the $y$ axis. 
As the transmitted signal propagates inside the plasma it leaves behind a trail of electrostatic disturbances as evident from the subplots of $E_x$ and  $J_x$ (the $x$ component of the electric field and current density respectively).
The unbalanced ion  (red color) and electron (blue) densities are also evident from the subplots of the Fig.\textcolor{blue}{\ref{fig:RLObervation}(f), (g) and (h)}. We would now like to characterize these disturbances 
(electromagnetic as well as electrostatic observed in the plasma). 

\begin{figure}
	\centering
	\includegraphics[width=6.2in]{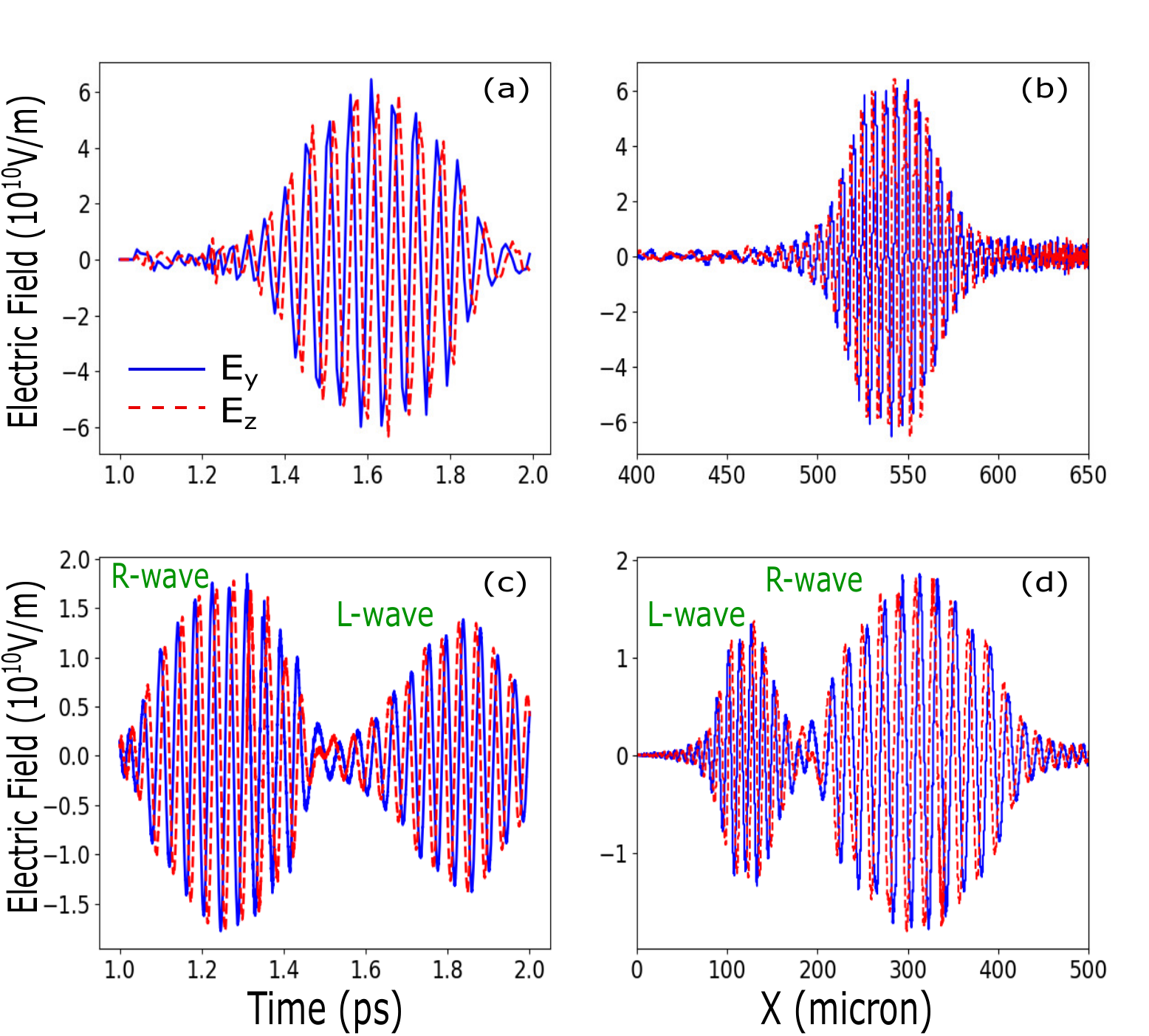}
	\caption{Phase variation of electric field $E_y$ and $E_z$ at $x=200 \mu m$ (a) for case (A) and (c) for case (B). Phase variation of electric field $E_y$ and $E_z$ at $t=2ps$ (b) for case (A) and (d) for case (B). It evident that for case (A) the excited mode is Right Circularly Polarized (R-mode) and for case (B) both Right and Left Circularly Polarized modes are obtained. Also (c) and (d) indicate that the R and L modes are separated in time and space.}
	\label{fig:WhistlerPolarization}
\end{figure}

\begin{figure}
	\centering
	\includegraphics[width=6.2in]{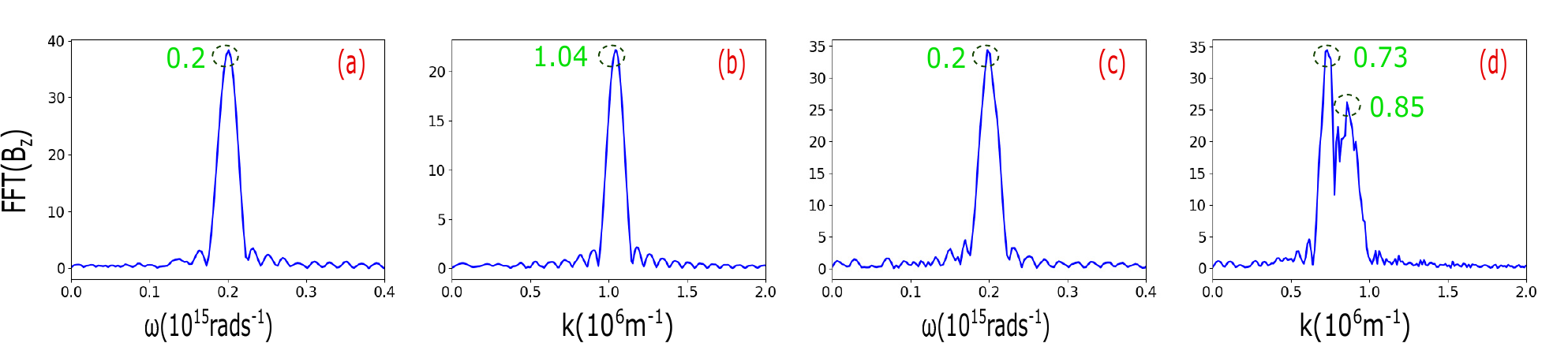}
	\caption{{(a) and (c) indicate the Fast Fourier Transform (FFT)  for electromagnetic field $B_z$ in time at $x=200\mu m$. ((b) and (d)) indicate the FFT for $B_z$ in space at $t=2ps$. FFT in (a) and (b) are for case (A). FFT in (c) and (d) are for case (B). The FFT is taken after the laser has reflected from the vacuum plasma interface.}}
	\label{fig:WhistlerFFTBz}
\end{figure}

We observe that though the incident laser was plane-polarized with its electric field component directed along $y$, the transmitted electromagnetic signal has both $y$ and $z$ components (Fig.\textcolor{blue}{\ref{fig:WhistlerPolarization}}). Furthermore, from the phase lag between the 
$E_y$ and $E_z$ components (Fig.\textcolor{blue}{\ref{fig:WhistlerPolarization} (a), (b)}) we can identify as the right circularly polarised EM radiation. 
The Fast Fourier Transform (FFT) of the $B_z$ signal in time at $x=200 \mu m$ and in space at $t=2ps$ has been shown in Fig.\textcolor{blue}{\ref{fig:WhistlerFFTBz} (a), (b)} respectively. The frequency peak occurs at the laser frequency $0.2 \times 10^{15} rad/sec = 0.2 \omega_{pe} $ and the  wavenumber from the spatial FFT  
is found to be  $k = 1.04 \times 10^6 m^{-1}$ (which is distinct from the laser wavenumber being $0.67\times10^6m^{-1}$).  

{The R and L-wave dispersion relations \cite{BoydSanderson} are as follows:}

\begin{equation}
	\label{Eq:Rwavedispersion}
	n^2 = \frac{c^2 k^2}{\omega^2} = 1 - \frac{\omega_{pe}^2}{\left(\omega + \omega_{ci} \right)\left(\omega - \omega_{ce} \right)}
\end{equation}

\begin{equation}
	\label{Eq:Lwavedispersion}
	n^2 = \frac{c^2 k^2}{\omega^2} = 1 - \frac{\omega_{pe}^2}{\left(\omega - \omega_{ci} \right)\left(\omega + \omega_{ce} \right)}
\end{equation}

\begin{figure}
	\centering
	\includegraphics[width=6.2in]{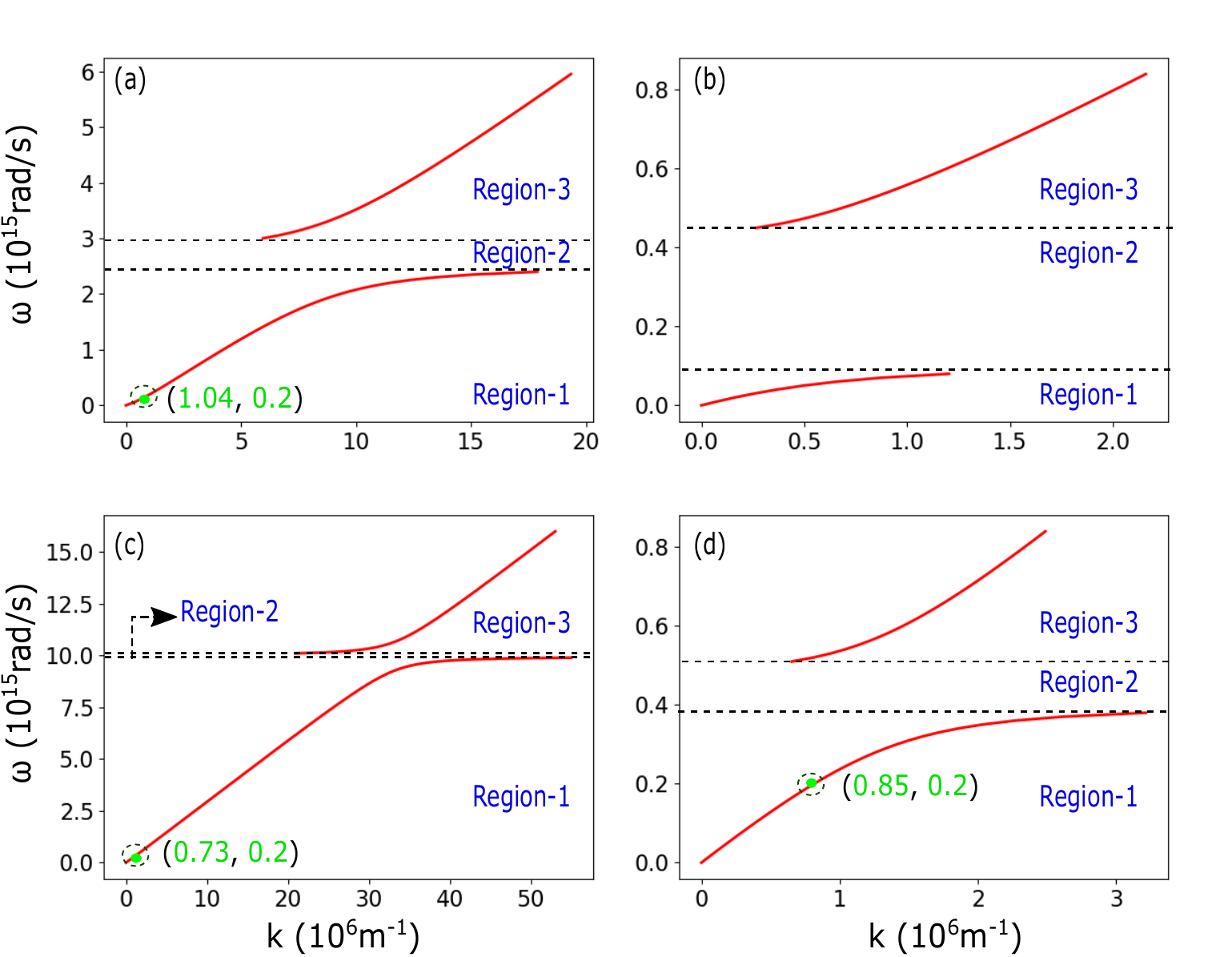}
	\caption{Diagram of dispersion curves for case (A) ((a) R-mode (a) and (b) L-mode) and for case (B) ((c) R-mode (c) and (d) L-mode). For R, L-mode geometry, region-1 and region-3 are passbands while region-2 is stopband. For case (A), electromagnetic wave frequency ($\omega = 0.2\times 10^{15} rad/s$) lies (a) in the passband of R-mode and (b) in the stop band of L-mode. For case (B), electromagnetic wave frequency ($\omega = 0.2\times 10^{15} rad/s$) lies in the passband of both R-mode and L-mode. For case (A), only R-mode is obtained, while for case (B), both R-mode and L-mode are obtained.}
	\label{fig:DispersionCurve}
\end{figure}

 {This frequency and wavenumber lies in the pass band for R-mode and stop band for L-mode Fig.\ref{fig:DispersionCurve} (a) and (b) respectively} and  satisfy the R wave dispersion relation Eq.(\textcolor{blue}{\ref{Eq:Rwavedispersion}}) of the EM wave inside the plasma. The EM disturbance in the plasma thus clearly gets identified as an R wave.

\begin{figure}
	\centering
	\includegraphics[width=6.2in]{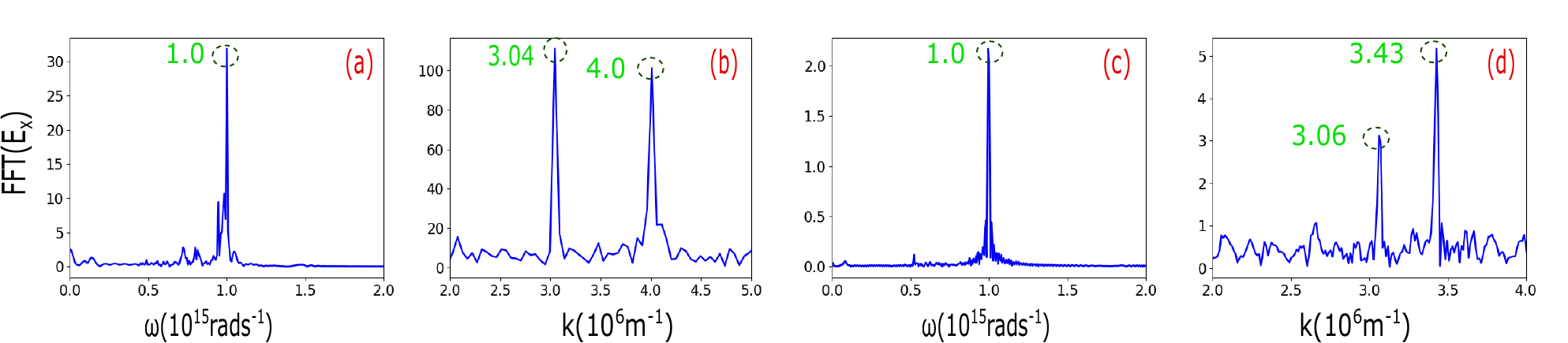}
	\caption{{(a) and (c) indicate the Fast Fourier Transform (FFT)  for electrostatic field $E_x$ in time at $x=200\mu m$. ((b) and (d)) indicate the FFT for $E_x$ in space at $t=2ps$. FFT in (a) and (b) are for case (A). FFT in (c) and (d) are for case (B). The FFT is taken after the laser has reflected from the vacuum plasma interface.}}
	\label{fig:WhistlerFFTEx}
\end{figure}

We now study the characteristic features of the electrostatic disturbance which gets created in the plasma. 
The FFT in time at $x = 200 \mu m$ and in space at $t = 2ps$ have been shown in Fig.\textcolor{blue}{\ref{fig:WhistlerFFTEx} (a), (b)} respectively. 
{For the chosen configuration ($k$ is parallel to the $B_0$), beside the R and L wave, there is also an elctrostatic mode present with a frequency same as that of electron plasma frequency. } The wavenumber, however,  shows two peaks at $3.04\times 10^6m^{-1}$ and $4.0\times 10^6m^{-1}$. {We shall understand the physics behind these observations in the next section.}

\subsection{Case (B)}
\begin{figure*}
	\centering
	\includegraphics[width=6.2in]{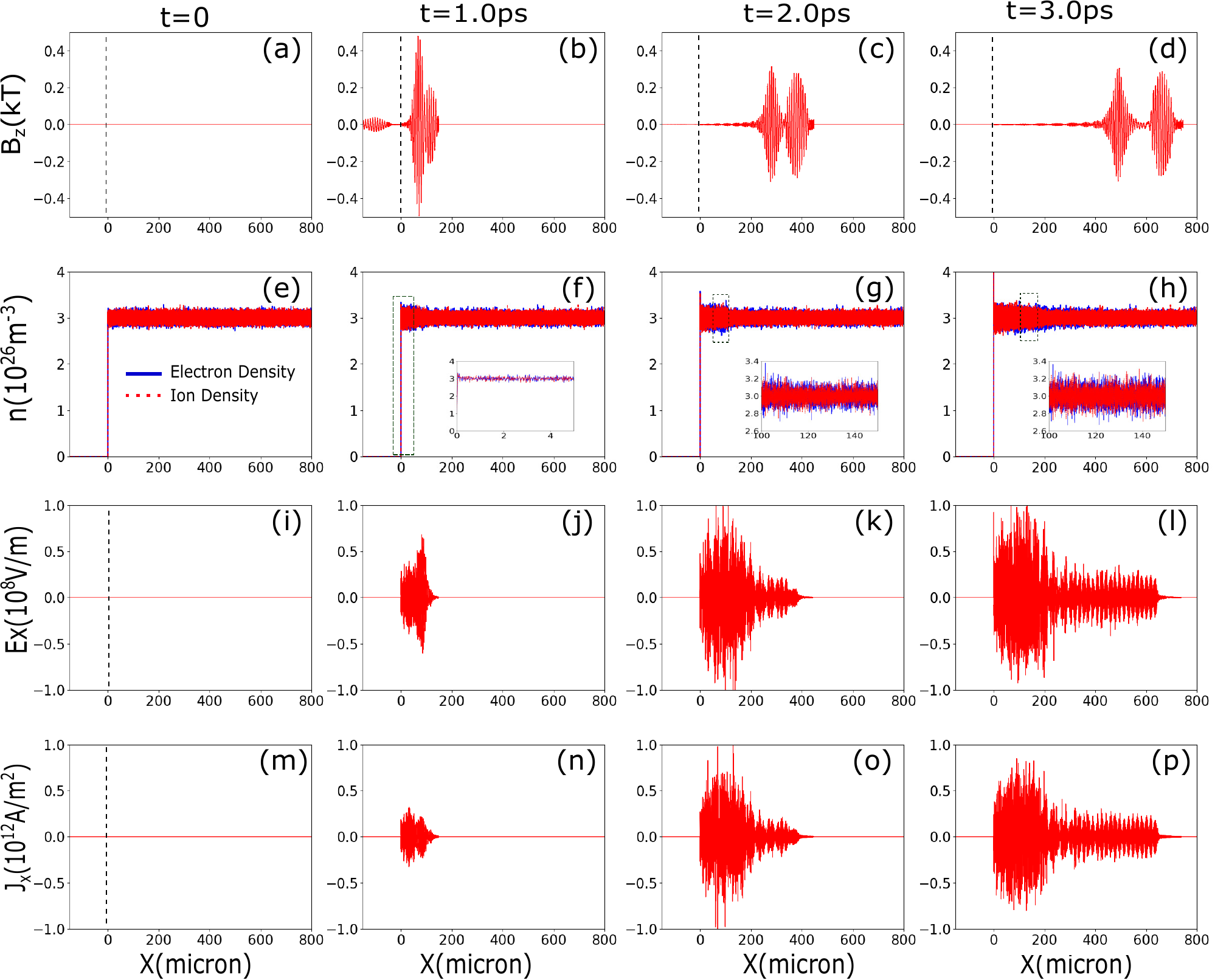}
	\caption{Time evolution of spatial variation of electromagnetic field $B_z$, density ($n$) perturbation of electron and ions, $x$-component of electric field $E_x$ and $x$-component of total current $J_x$ with respect to $\hat{x}$ for case (B), showing that as the laser interacts with magnetized plasma, there is finite magnitude of perturbed $B_z$, $E_x$ and $J_x$. We can infer from these plots that a part of laser energy has been coupled into plasma.}
	\label{fig:RLObervation}
\end{figure*}

We now consider  the laser frequency to satisfy  the condition ($\omega_{l}<\omega_{ci}<\omega_{ce}$). For this case, both electrons and ions would exhibit a magnetized response at the laser frequency.  In Fig.\textcolor{blue}{\ref{fig:RLObervation}} we show the evolution of various fields as a function of $x$ at different times (viz.,$t=0$, $t=1.0ps$, $t=2.0ps$ and $t=3.0ps$) for this particular case. In contrast to case(A), in this case, the transmitted magnetic field of the EM wave shows two distinct pulses. The spatial separation between the two pulses is observed to increase with time.  In this case too 
as the transmitted EM signal propagates inside the plasma it leaves behind a trail of electrostatic disturbances as evident from the subplots of $E_x$ and $J_x$ (the $x$ component of the electric field and current density respectively).

To check the polarization of the transmitted EM signal we 
plot the $E_y$ and $E_z$ component of the electric field at a particular $x$ location ($x=200 \mu m$) as a function of time in  Fig.\textcolor{blue}{\ref{fig:WhistlerPolarization}(c)} {and at particular time ($t=2ps$) as a function of space in  Fig.\textcolor{blue}{\ref{fig:WhistlerPolarization}(d)}}.  It is clear from the plot that the pulse which moves ahead has a right-hand circular polarization whereas the trailing pulse has the Left-hand circular polarization. 
The FFT of the $B_z$ signal in space and time has been shown in Fig.\textcolor{blue}{\ref{fig:WhistlerFFTBz}(c), (d)} respectively. The frequency peak occurs at the laser frequency $0.2\times10^{15} rad/sec = 0.2\omega_{pe}$ {which lies in the passband of R-mode and L-mode as evident from Fig.\ref{fig:DispersionCurve} (c) and (d) respectively} and the wavenumbers from the FFT are found to be $k_1=0.73\times10^{6} m^{-1}$ and $k_2=0.85\times10^6 m^{-1}$.  This frequency and wavenumbers ($k_1$ and $k_2$)  satisfy the R wave and L wave dispersion relation Eq.(\textcolor{blue}{\ref{Eq:Rwavedispersion}}) and Eq.(\textcolor{blue}{\ref{Eq:Lwavedispersion}}) respectively of the EM wave inside the magnetized plasma.   

The FFT in space at $t=2ps$ and time at $x=200 \mu m$ has been shown in Fig.\textcolor{blue}{\ref{fig:WhistlerFFTEx} (c), (d)}  for the  
electrostatic oscillations that get generated inside the plasma. 
In this case too the 
frequency of the electrostatic disturbances is observed to match with the electron plasma frequency. {However, two wavenumber peaks are obtained at $3.06\times 10^6 m^{-1}$ and $3.43\times 10^6 m^{-1}$. We shall discuss the physics behind these observation in the next section.}    

\section{The genesis of electrostatic excitation}
\label{ElectrostaticExcitation}

It should be noted that for a  pulsed EM field, a ponderomotive force, proportional to the gradient of EM wave intensity is operative on the plasma particles. In our 1-D problem with pulsed EM field, this force will be operative along the direction of propagation of the EM pulse, and the corresponding acceleration felt by the two species is given by  the following expression: 

\begin{equation}
	\label{ponderomotive}
	{{{\frac{\partial {u^{e,i}_x}}{\partial t}}  = -\frac{e^2 (1 +\alpha^2 \pm 2 \alpha \omega_{c{e,i}}/\omega_l)}{m_{e,i}^2\omega_l^2\left(1-\omega^2_{c{e,i}}/\omega_l^2 \right)}\nabla E^2(x)}}
\end{equation}   
Here, the plus sign is for electrons, and the minus sign is for ions. 
Thus, $\omega_{ce} = eB_0/m_e$ and $\omega_{ci} = eB_0/m_i$ are the magnitude of the   electron and ion cyclotron frequencies respectively. {Where $m_e$ and $m_i$ are respectively the electron and ion mass.} Also, $\omega_{l}$ stands for the laser frequency and $e$ represent the magnitude of electronic charge. Here $\alpha$ takes the value of $0$ and $\pm 1$ for plane and right and left circularly  polarised EM (electromagnetic) waves respectively.  The derivation of  Eq.(\textcolor{blue}{\ref{ponderomotive}}) has been provided  in \ref{Append:RL}.  
In the absence of an applied magnetic field, the above equation reduces to the standard form of {(as given in \ref{Append:unmagnetized}})
\begin{equation}
	\label{unmagnetized}
	{{{\frac{\partial {u^{e,i}_x}}{\partial t}}  = -\frac{e^2 (1 +\alpha^2 )}{m_{e,i}^2\omega_l^2}\nabla E^2(x)}}
\end{equation}
which shows that the electrons experience a stronger ponderomotive acceleration. 

The physics of the electrostatic fluctuation driven by the difference in the ponderomotive force experienced by electrons and ions can be understood as follows. From Eq.(\textcolor{blue}{\ref{ponderomotive}}) it is clear that the electrons and ions acquire an additional velocity (other than the quiver velocity in the laser field) along $x$ direction. The difference between the $x$ component of ion and electron velocity generates $x$ component of current. The laser propagation direction being along $x$, this current has a finite divergence.  
From the charge continuity equation, viz., 
\begin{equation}
	\frac{\partial \rho}{\partial t} + \vec{\nabla}\cdot\vec{J} = 0
\end{equation} 
the current divergence leads to space charge fluctuations which have been observed in our simulations (second row of Fig.\textcolor{blue}{\ref{fig:WhistlerObervation}} and Fig.\textcolor{blue}{\ref{fig:RLObervation}}). We have thus identified a new ponderomotive force-driven mechanism of electrostatic wave excitation in the plasma. 

In the limit of strong magnetic field for which $\omega_l << \omega_{ci} << \omega_{ce}$ (e.g. Case (B) of our simulations),  
Eq.({\textcolor{blue}{\ref{ponderomotive}}}) reduces for the two species to  
\begin{equation}
	\label{highB}
	{{{\frac{\partial {u^{e,i}_x}}{\partial t}}  = \frac{(1 +\alpha^2 \pm 2 \alpha  \omega_{c{e,i}}/\omega_l)}{B_0^2}\nabla E^2(x)}}
\end{equation}
Clearly, in this limit of the strong 
magnetic field the acceleration of the two species depends predominantly on the applied magnetic field and reduces as $ \sim 1/B_0^2$. 
The two species experience different acceleration only for circularly polarised waves for which $\alpha = \pm 1$. 
For plane polarization with $\alpha = 0$ both the species 
experience identical ponderomotive acceleration. In this case thus one would not generate any charge fluctuation.  
When the laser frequency falls in the regime of $\omega_{ci} << \omega_l << \omega_{ce}$ (case (A) of our simulations) the ion would experience the unmagnetized ponderomotive expression provided by Eq.(\textcolor{blue}{\ref{unmagnetized}}) and electron 
would be governed by the Eq.(\textcolor{blue}{\ref{highB}}). 

\begin{figure*}
	\centering
	\includegraphics[width=6.2in]{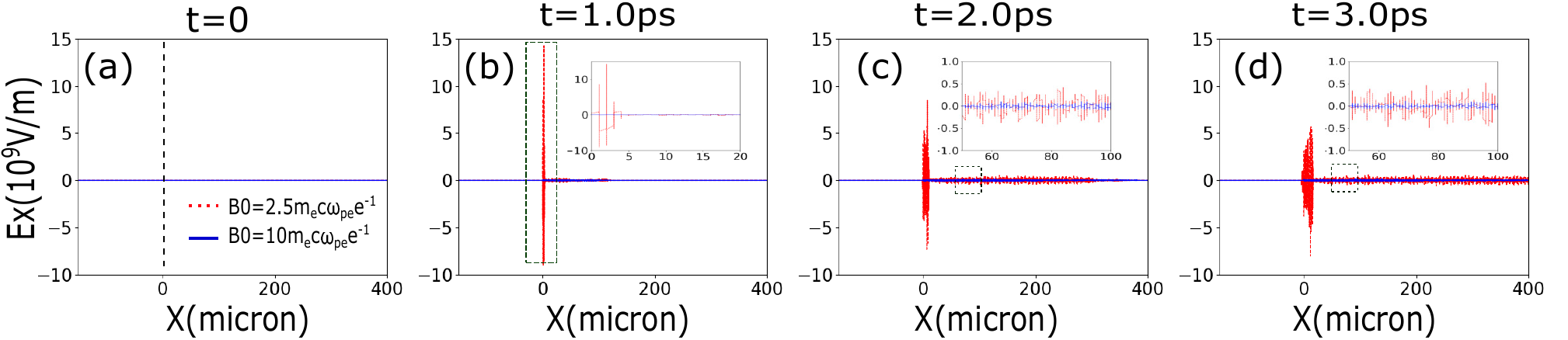}
	\caption{Time evolution of Electrostatic field variation for case (A) (red desh line) and case (B) (blue line). It is evident from the plots that the magnitude of electrostatic field generation due to ponderomotive force is more in case (A) than in case (B) }
	\label{fig:ExVariation}
\end{figure*}

Let us now try to analyze our observations from these considerations.  For case (A) since the ion and electron satisfy the unmagnetized and magnetized expressions for the ponderomotive acceleration, for this case, there will always be an unbalanced current (for the plane as well as circular polarization). { At the plasma vacuum surface the ponderomotive pressure of the plane incident laser is operative and electrostatic fluctuations are evident as clear from the zoomed view of the electrostatic field in Fig.\textcolor{blue}{\ref{fig:WhistlerObervation}}}.   This should be contrasted with case (B) where there are no surface electrostatic excitations. Both ion and electrons here satisfy Eq.(\textcolor{blue}{\ref{highB}}) which is identical for plane polarization of the incident laser pulse. The overall magnitude of the electrostatic fluctuations in case (B) typically generated by the L and R mode in the bulk of the plasma is also comparatively much low compared with case(A) {as evident from Fig.\textcolor{blue}{\ref{fig:ExVariation}}}. 

\begin{figure}
	\centering
	\includegraphics[width=6.2in]{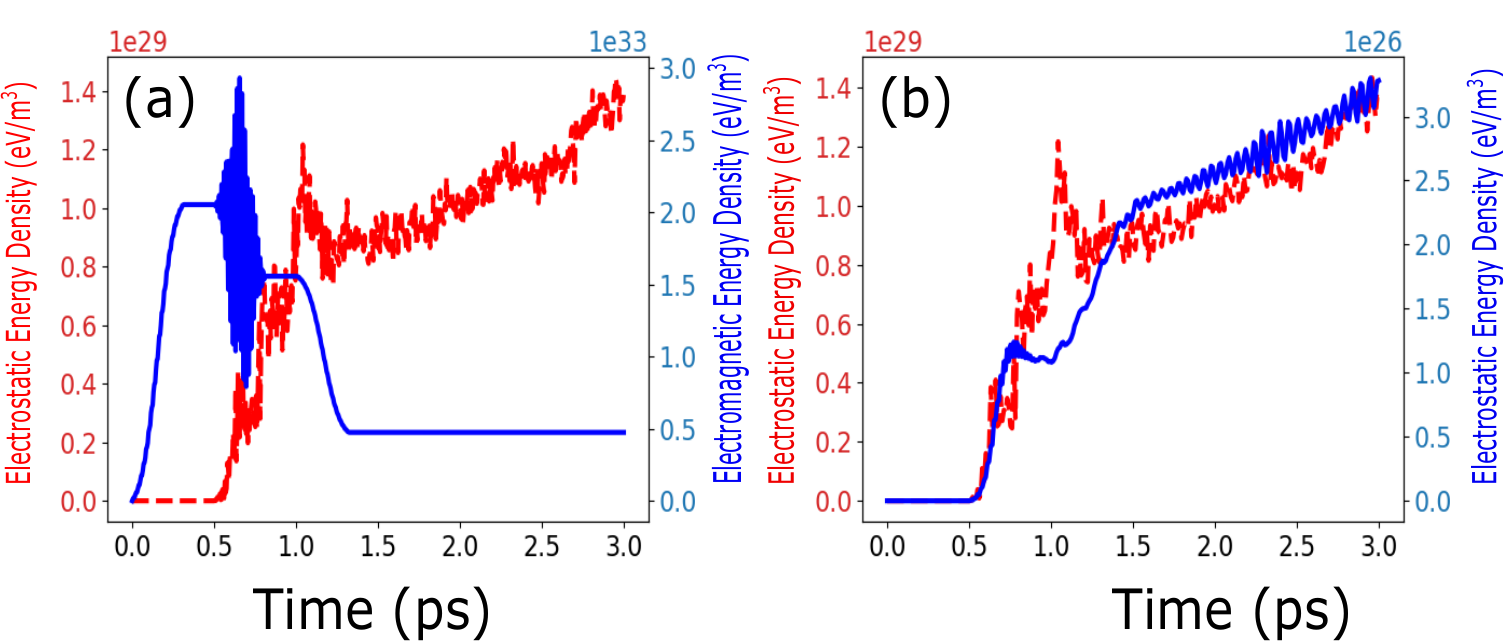}
	\caption{The comparision for the time variation of (a) for case (A) the mean of Electrostatic field energy density (red desh line, left $y$ axis) and electromagnetic (blue line, right $y$ axis), and (b) the mean of Electrostatic field energy density for case (A) (red desh line, left $y$ axis) and case (B) (blue line, right $y$ axis). It is evident from subplot (a) that electrostatic modes are generated along with the electromagnetic field. Subplot (b) signifies that the electrostatic field eenrgy density reduces with increase in external magnetic field and electrostatic field energy is more in case(A) than that of in case (B).}
	\label{fig:EnergyB25B10}
\end{figure}

\begin{figure}
	\centering
	\includegraphics[width=4in]{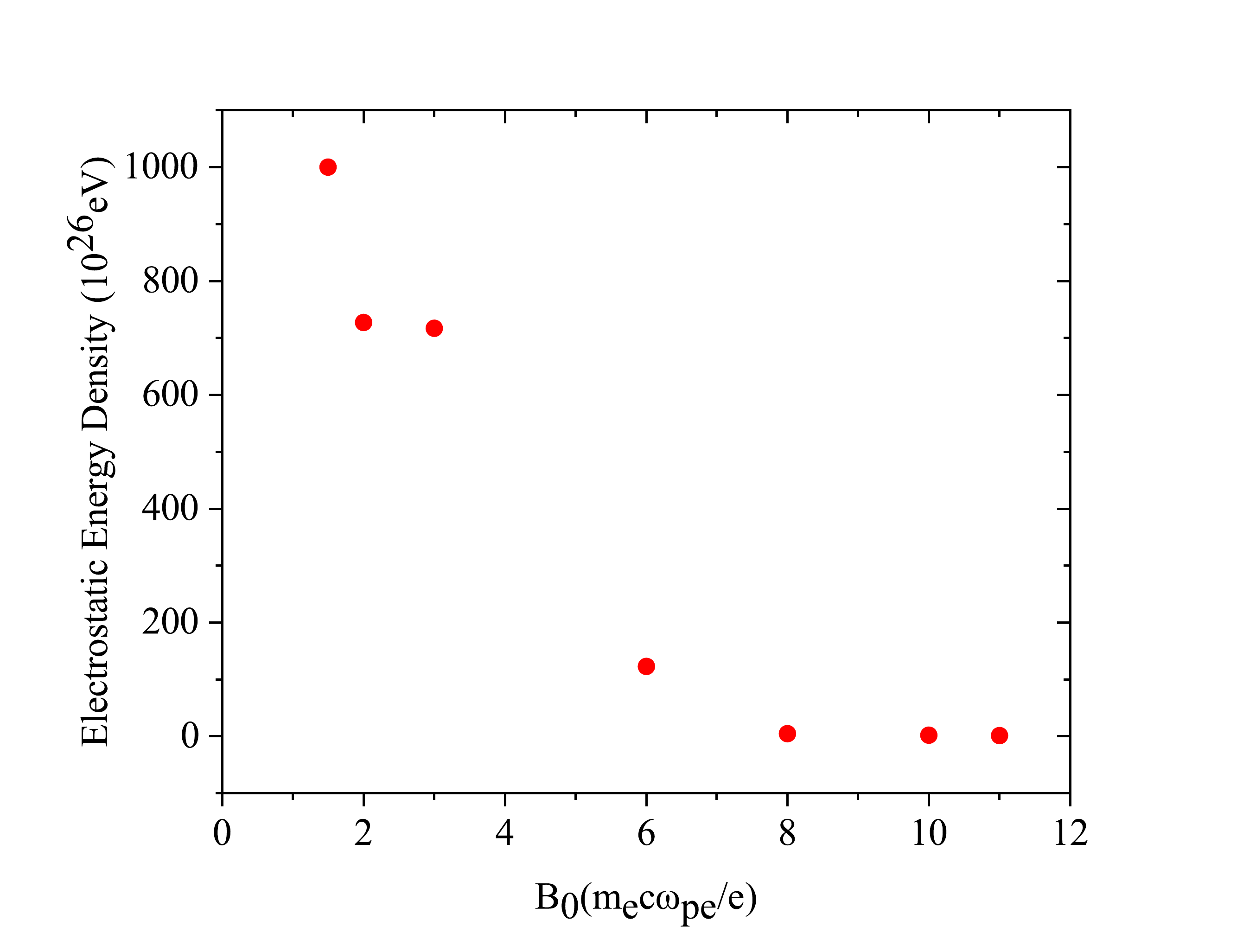}
	\caption{Figure shows the effect of varying external magnetic field on the Electrostatic Energy. It can be observed from the graph that the energy absorbed in the plasma medium reduces predominantly at higher magnetic fields.}
	\label{fig:ElectrostaticEnergyvsB0}
\end{figure}

{Energy density plot in Fig.\ref{fig:EnergyB25B10}(a) signifies the generation of electrostatic field energy for case (A). It is also evident from Fig.\ref{fig:EnergyB25B10}(b) that the eelctrostatic field energy generation is more in case (A) than in case(B).} We have repeated our simulations with the increasing magnitude of the magnetic field.   In Fig.\textcolor{blue}{\ref{fig:ElectrostaticEnergyvsB0}} we provide a comparison of the evolution of the electrostatic field energy in the plasma for different values of $B_0$. 
It can be seen clearly, that the 
electrostatic energy in the plasma reduces with the increasing magnetic field.


These studies thus show that the observations on electrostatic excitations and absorption are consistent 
with the proposed mechanism driven by the difference between the ponderomotive acceleration felt by the two species.


\normalsize

\section{Conclusion}
\label{conclusion}
We have carried out extensive PIC simulations with the help of  EPOCH-4.17.10 for the propagation of laser pulse in a magnetized plasma. 
The direction of the external magnetic field has been chosen along the laser propagation direction. We demonstrate a novel mechanism of electrostatic wave excitation leading to laser energy absorption in the plasma medium. 
The mechanism is essentially driven by the difference between the ponderomotive acceleration felt by electrons and ions.

\section{Acknowledgment} 

The authors would like to acknowledge the EPOCH Consortium, for providing us the access to the EPOCH-4.17.10 framework \textcolor{blue}{\cite{EPOCH}}. This research work has been supported by the Core Research Grant No. CRG/2018/000624 of Department of Scient and Technology (DST), Government of India. We also acknowledge support from J. C. Bose Fellowship grant of AD (JCB-000055/2017) from the Science and Engineering Research Board (SERB), Government
of India. The authors thank IIT
Delhi HPC facility for computational resources. L.P.G also wishes to thank Sathi Das for discussions. 

\section{References}
\bibliographystyle{unsrt}
\bibliography{PlasmaPhysicsandControlledFusion.bib}

\newpage
\appendix
\section{General derivation for Ponderomotive Force in magnetized plasma for elliptically polarized laser}
\label{Append:GeneralDerivation}

To derive for the ponderomotive force, we consider the response of a homogeneous magnetized plasma to a high frequency elliptically polarised field whose amplitude is spatially varying along $\hat{x}$ given as:

\begin{equation}
	\vec{E} = E(x) \left(\hat{y} + i\alpha\hat{z}\right)e^{-i\omega t}
\end{equation}

Using the Maxwell equation 
\begin{equation}
	\nabla \times \vec{E} = -\frac{\partial \vec{B}}{\partial t}
\end{equation}

We calculate the oscillating magnetic field as:
\begin{equation}
	\label{Eq:Magnetic}
	\vec{B}_1 = -\frac{i}{\omega}\frac{\partial E(x)}{\partial x}\left(\hat{z} - i\alpha\hat{y}\right)e^{-i\omega t}
\end{equation}

We are considering a constant external magnetic field given by the following expression: 

\begin{equation}
	\vec{B}_{ext} = B_x\hat{x} + B_y\hat{y} + B_z\hat{z} = b_xB_0\hat{x} + b_yB_0\hat{y} + b_zB_0\hat{z}
\end{equation}

Where, 
\[ b_i = \frac{B_i}{B_0} \]

Thus the total magnetic field is summation of time varying oscillating magnetic field and the constant external magnetic field given as:

\begin{equation}
	\vec{B} = \vec{B}_{ext} + \vec{B}_1
\end{equation}

\begin{figure*}
	\centering
	\includegraphics[width=6in]{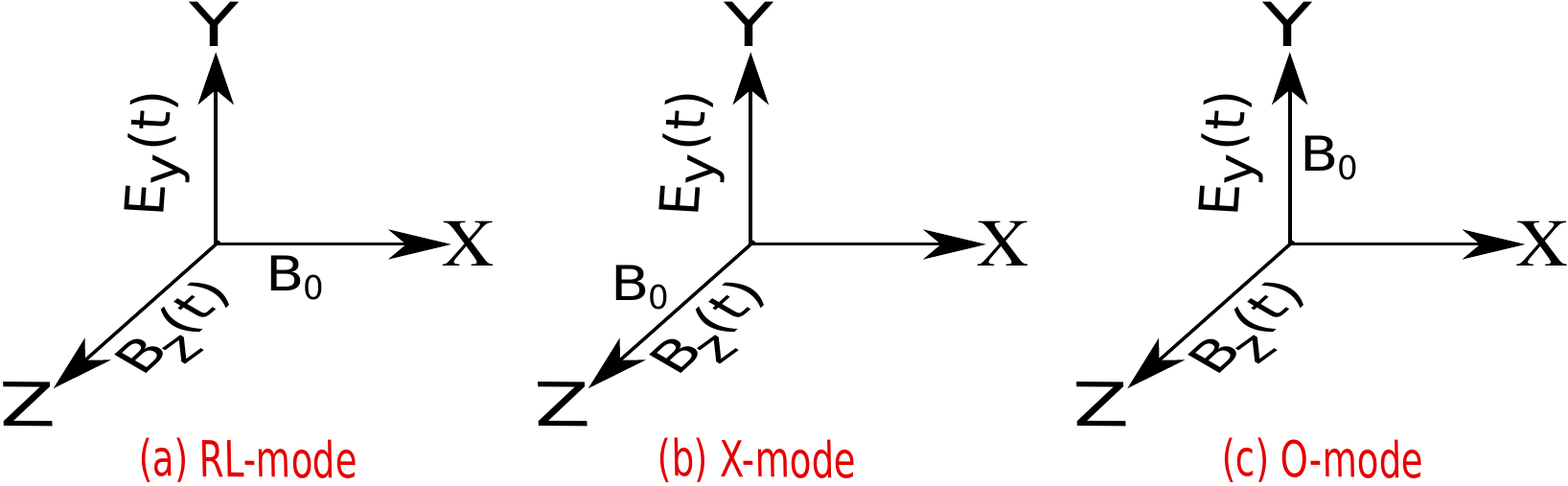}
	\caption{Geometry for (a) RL mode (the external magnetic field $B_0$ is applied along laser propagation direction $\hat{x}$ with oscillating laser electric field along $\hat{y}$ direction), (b) X-mode (the external magnetic field $B_0$ is applied along direction of laser magnetic field $\hat{z}$ with oscillating laser electric field along $\hat{y}$ direction), and (c) O-mode configurations (the external magnetic field $B_0$ is applied along direction of laser electric field $\hat{y}$).}
	\label{fig:RLgeometry}
\end{figure*}


If we neglect the electron - ion pressure terms, the force equation is given as:

\begin{equation}
	\frac{\partial\vec{u}}{\partial t} + \vec{u}\cdot\nabla\vec{u}  = \frac{q_p}{m_p}\left(\vec{E} + \vec{u}\times\vec{B}\right)
\end{equation}

Where $p = e, i$ respectively indicate the electron and ion species. So $q_e$ and $q_i$ are the electron and ion charges. Similarly $m_e$ and $m_i$ are the electron and ion masses. 

To lowest order in $|\vec{E}|$, $\vec{u} = \vec{u}^h$ we can ignore the non-linear terms from the momentum equation 

\begin{equation}
	\label{Eq:higherFreq}
	\frac{\partial\vec{u}^h}{\partial t}  = \frac{q_p}{m_p}\left(\vec{E} + \vec{u}^h\times\vec{B}{_2}\right)
\end{equation}

After taking fourier transform of the above Eq.(\ref{Eq:higherFreq}), and solving for $u_x^h$, $u_y^h$ and $u_z^h$ 

\begin{equation}
	\label{Eq:xf}
	u_x^h = i\frac{\omega_{cp}v_E}{\beta\omega^2}\left[\left(b_y\alpha - b_xb_y\frac{\omega_{cp}}{\omega}\right) + i\left(b_z - \alpha b_xb_z\frac{\omega_{cp}}{\omega}\right)\right]
\end{equation}

\begin{equation}
	\label{Eq:yf}
	u_y^h = \frac{v_E}{\beta\omega}\left[\alpha b_yb_z \frac{\omega^2_{cp}}{\omega^2} + i\left(1 - b_y^2\frac{\omega^2_{cp}}{\omega^2} - b_x\alpha\frac{\omega_{cp}}{\omega}\right)\right]
\end{equation}

\begin{equation}
	\label{Eq:zf}
	u_z^h = -\frac{v_E}{\beta\omega}\left[\alpha \left(1 - b_z^2\frac{\omega^2_{cp}}{\omega^2}\right) - b_x\frac{\omega_{cp}}{\omega} + ib_yc_z\frac{\omega^2_{cp}}{\omega^2}\right]
\end{equation}

Where, 

\begin{equation}
	v_E = \frac{q_p E(x)}{m_p}e^{-i\omega t}
\end{equation}

\begin{equation}
	\omega_{cp} = \frac{q_pB_0}{m_p}
\end{equation}

\begin{equation}
	\label{Eq:beta}
	\beta = 1 - \frac{\omega_{cp}^2}{\omega^2}\left(b_x^2 + b_y^2 + b_z^2\right) = 1 - \frac{\omega_{cp}^2}{\omega^2}\frac{B_{ext}^2}{B_0^2}
\end{equation}



The electrons and ions are simply oscillating in the oscillating electric and magnetic fields. By averaging the force equation over these rapid oscillations, we obtain

\begin{equation}
	\label{Eq:smallMomentum}
	\frac{\partial \vec{u}^s}{\partial t} = \frac{q_p}{m_p} \left<\vec{u}^h\times\vec{B}_1\right>_t - \left<\vec{u}\cdot\nabla\vec{u}\right>_t
\end{equation}

Where $\left< \right>_t$ denotes an average over high frequency oscillations. The second term in Eq.(\ref{Eq:smallMomentum}) can be written as: 





Substituting the known values in above equation, we have:

\begin{equation}
	\label{Eq:magneticForce}
	\frac{q_p}{m_p} \left<\vec{u}^h\times\vec{B}_1\right>_t = -\frac{q_p^2 }{2m_p^2 \beta \omega^2} \left[\left(1 - b_y^2 \frac{\omega_{cp}^2}{\omega^2}\right) + \alpha^2\left(1 - b_z^2 \frac{\omega_{cp}^2}{\omega^2}\right) - 2\alpha b_x \frac{\omega_{cp}}{\omega}\right]\nabla |E|^2
\end{equation}

\begin{equation}
	\label{Eq:convetive}
	\left<\vec{u}\cdot\nabla\vec{u}\right>_t = \frac{q^2_p \omega^2_{cp}}{2m^2_p \beta^2 \omega^4} \left[\left(\alpha b_y - b_xb_y \frac{\omega_{cp}}{\omega}\right)^2 + \left(b_z - \alpha b_xb_z \frac{\omega_{cp}}{\omega}\right)^2\right]\nabla|E|^2
\end{equation}

Using Eq.(\ref{Eq:magneticForce}) and Eq.(\ref{Eq:convetive}), we can write Eq.(\ref{Eq:smallMomentum}) as:


\begin{equation}
	\label{Eq:smallMomentumfinal}
	\frac{\partial \vec{u}^s}{\partial t} = -\frac{q_p^2 \nabla |E|^2 }{2m_p^2 \beta \omega^2}\left[A_1 + \alpha^2A_2 - 2\alpha a \frac{\omega_{cp}}{\omega}  + \frac{\omega_{cp}^2}{\beta\omega^2}A_3\right]
\end{equation}


Where, 
\[ A_1 = \left(1 - b_y^2 \frac{\omega_{cp}^2}{\omega^2}\right) \]
\[ A_2 = \left(1 - b_z^2 \frac{\omega_{cp}^2}{\omega^2}\right) \] 
\[ A_3 = \left(\alpha b_y - b_xb_y \frac{\omega_{cp}}{\omega}\right)^2 + \left(b_z - \alpha b_xb_z \frac{\omega_{cp}}{\omega}\right)^2 \]

\subsection{\textcolor{blue}{Un-magnetized case}}
\label{Append:unmagnetized}
For unmagnetized case, we have $b_x = b_y = b_z = 0$, thus $\beta = 1$. So the ponderomotive force in this case will be \cite{nishikawa}: 

\begin{equation}
	\label{Eq:unmagnetized}
	\frac{\partial \vec{u}^s}{\partial t} = -\frac{q_p^2 \nabla |E|^2 }{2m_p^2 \omega^2}\left(1 + \alpha^2\right)
\end{equation}

\subsection{\textcolor{blue}{RL-mode geometry}}
\label{Append:RL}
For RL-mode case, we have $b_x = 1$ and $b_y = b_z = 0$, thus $\beta = 1 - \frac{\omega_{cp}^2}{\omega^2}$. So the ponderomotive force in this geomrety is will be: 

\begin{equation}
	\label{Eq:RLcase}
	\frac{\partial \vec{u}^s}{\partial t} = -\frac{q_p^2 \nabla |E|^2 }{2m_p^2 \omega^2}\frac{\left[1 + \alpha^2 - 2\alpha \frac{\omega_{cp}}{\omega}\right]}{\left(1 - \frac{\omega_{cp}^2}{\omega^2}\right)}
\end{equation}

\subsection{\textcolor{blue}{X-mode geometry}}
\label{Append:X}
For X-mode case, we have $b_x = b_y = 0$ and $b_z = 1$, thus $\beta = 1 - \frac{\omega_{cp}^2}{\omega^2}$. So the ponderomotive force in this geomrety is will be: 

\begin{equation}
	\label{Eq:Xmodecase}
	\frac{\partial \vec{u}^s}{\partial t} = -\frac{q_p^2 \nabla |E|^2 }{2m_p^2 \omega^2}\frac{1}{\left(1 - \frac{\omega_{cp}^2}{\omega^2}\right)}\left[1 + \alpha^2\left(1 - \frac{\omega_{cp}^2}{\omega^2}\right) + \frac{\omega_{cp}^2}{\omega^2}\frac{1}{\left(1 - \frac{\omega_{cp}^2}{\omega^2}\right)}\right]
\end{equation}

\subsection{\textcolor{blue}{O-mode geometry}}
\label{Append:O}
For O-mode case, we have $b_x = b_z = 0$ and $b_y = 1$, thus $\beta = 1 - \frac{\omega_{cp}^2}{\omega^2}$. So the ponderomotive force in this geomrety is will be: 

\begin{equation}
	\label{Eq:Omodecase}
	\frac{\partial \vec{u}^s}{\partial t} = -\frac{q_p^2 \nabla |E|^2 }{2m_p^2 \omega^2}\frac{1}{\left(1 - \frac{\omega_{cp}^2}{\omega^2}\right)}\left[\left(1 - \frac{\omega_{cp}^2}{\omega^2}\right) + \alpha^2 + \frac{\omega_{cp}^2}{\omega^2}\frac{\alpha^2}{\left(1 - \frac{\omega_{cp}^2}{\omega^2}\right)}\right]
\end{equation}


\end{document}